# Energy recovery operation for continuous-wave X-ray free-electron lasers

Guanglei Wang[1], Jiawei Yan[2,3], Nanshun Huang[2,3], Duan Gu[4], Meng Zhang[4], Haixiao Deng[4*], Bo Liu[4], Dong Wang[4], Xueming Yang[1] and Zhentang Zhao[4]

[1]State Key Laboratory of Molecular Reaction Dynamics, Dalian Institute of Chemical Physics, Chinese Academy of Sciences, Dalian 116023, P. R. China

[2]Shanghai Institute of Applied Physics, Chinese Academy of Sciences, Shanghai 201800, P. R. China

[3]University of Chinese Academy of Sciences, Beijing 100049, P. R. China

[4]Shanghai Advanced Research Institute, Chinese Academy of Sciences, Shanghai 201800, P. R. China

*Author to whom correspondence should be addressed: denghaixiao@zjlab.org.cn

**Abstract**: A superconducting linear accelerator operating in continuous-wave mode could produce X-ray free electron lasers (XFEL) at megahertz repetition rate, with the capability that delivering wide spectral range coherent radiation to multi end stations. In this Letter, the energy recovery Linac (ERL) mode is proposed to flexibly control the electron beam energy for a continuous-wave superconducting Linac. Theoretical investigations and multi-dimensional numerical simulations are applied to the Linac case of Shanghai high-repetition-rate XFEL and extreme light facility. The results show that, with ERL operation in the last 25 cryo-modules, the strict requirements on RF power system could be significantly relaxed. And if one exhaust the RF power, the maximum electron beam energy can be enhanced from 8.74 GeV to 11.41GeV in ERL mode. The optimization of the ERL operation, the multi-energy electron beam transport and the XFEL performance improvements are presented.

In recent years, enormous progresses have been achieved in the free-electron lasers (FELs) based on superconducting radio-frequency (SRF) Linac, which hold great potential to deliver unprecedented average power and high brilliance radiation pulses previously inaccessible by either storage ring light sources or XFELs driven by copper Linac. The feasibility of superconducting FEL user facility has already been fully demonstrated at TESLA in 2000 [1] and is currently used to deliver coherent FEL radiation from extreme ultraviolet to soft X-ray range to users routinely at FLASH.[2] In the following two decades, to meet the increasingly scientific needs in charactering matter on more precise atomic length and time scales, several hard X-ray superconducting FEL facilities are gradually coming out, i.e., European XFEL has achieved its first lasing and started user operation in 2017.[3] The Linac Coherent Light Source II, and the Shanghai high-repetition-rate XFEL and extreme light facility (SHINE) is currently under construction, and is scheduled to produce the first light in 2022 and 2025, respectively.[4,5]

Wavelength tunability is one of the most distinguishing features of an FEL light source, which enables the multiple user operation or element-specific spectroscopy experiments simultaneously.[2,6-11] According to the resonance condition, FEL central wavelength $\lambda$ can be presented as $\lambda = \lambda_u(1 + a_u^2)/2\gamma^2$, where the $\lambda_u$ is the undulator period length, $a_u$ is the undulator strength parameter, and $\gamma$ is the mean

Lorentz factor of the electron beam.[12] Even though the FEL wavelength can be controlled by adjusting the undulator field parameters, the FEL tuning range is still obviously limited by the performance of the undulator magnetic field.[13] Then changing electron beam energy is another more straightforward and efficient way to realize the tunable FEL wavelength, and several schemes has been developed and utilized in Linac operation to generate various electron beam energies.[9,14] For example, one can extract the electron beam with a fast kicker at a specific position of Linac and transport the beam to undulator by a bypass line. It is a general solution in current FEL facilities, in which not only the extracted beam energy will be limited by the capacity of the upstream Linac, but also an additional and expensive long beam transport bypass line is required. Moreover, in those schemes, the FEL wavelength tunability can be realized in a relatively large spectrum range, while the capability of continuously adjust the radiation wavelength is still limited practically.

Recently, two novel concepts have been proposed for multi-energy operation in a continuous-wave Linac, e.g., by using an achromatic and isochronous magnetic delay system, and by the off-frequency detuning of superconducting cavity to control the acceleration phase of the electron beam.[15,16] Those methods could generate electron beam with the energy approximately from zero up to the maximum energy capacity of the SRF Linac, and then extend the tuning range and performance of continuous-wave FEL facilities very certainly. In this Letter, on the basis of acceleration phase adjustment,[15] we explore the possibility of energy recovery Linac (ERL)[17,18] operation for a continuous-wave XFEL to generate electron beams which could even exceed the upper bound performance of a settled Linac parameter. As well known, the electro-magnetic field in microwave cavities can accelerate as well as decelerate electron beam. In an ERL, the electron beam is injected into the cavities at proper phase for acceleration, and then a return loop containing undulators and other beam instruments leads it form the end of the Linac back to its beginning, the electron beam traverses the same Linac for deceleration and the remaining energy of the electron beam is recovered into the electro-magnetic field again. The most attractive characteristic of ERLs is saving budget on RF drive while delivering extremely bright, high power electron beams, and significantly reduce the risk of radiation hazard at the dump simultaneously. The principle of ERL has been fully demonstrated at Stanford SCA/FEL in a 1985,[18] now is successfully developed and operated as FELs and some other kind of photon sources in several laboratories.[20-24]

The schematic layout of the ERL operation for continuous-wave XFEL is shown in Fig.1 and a set of parameters of SHINE is utilized in the discussion hereafter. SHINE is the first hard X-ray FEL in China, aiming at generating coherent FEL radiation from 0.4 to 25 keV with a repetition rate up to 1MHz.[5] As mentioned in Ref 15, a bunch-to-bunch switching system consists of four horizontal double bend achromat (DBA) is designed in the break section between the L3 and L4 for delaying electron beams. When the electron beam is delayed by a half of the 1.3GHz SRF period to the negative phase, it'll be decelerated in L4. As shown in Fig.1, the kinetic energy of those green electron beams will be transferred back to the electro-magnetic field again in L4, and is available for the acceleration of those red electron beams subsequently. Due to the net beam loading can be considered as zero in ERL operation, the L4 section could accelerate electron beam at very high gradient with only modest amounts of RF power. Then with the existed RF station and cryogenic system of SHINE Linac, we can change the output beam energy arbitrarily, and the maximum electron beam energy could also be boosted from 8.74GeV to about 11.41 GeV with the ERL operation. In following, the requirements for RF system in different operation mode, the working point optimization of the ERL operation, the feasibility of co-transportation of 0.5GeV and 11.41 GeV electron beams in L4, and FEL performance improvements of the 12.94 keV photon energy are presented.

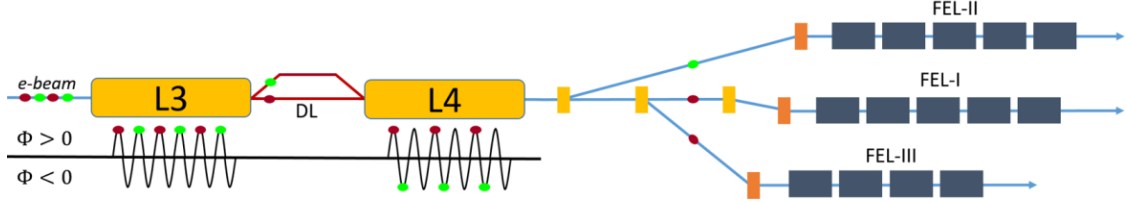

Fig. 1. The schematic layout of L3 and L4 section in SHINE Linac and the undulator configuration, the electron beam before L3 is about 2.1 GeV with a peak current of 2 kA, the DL section could be utilized to delay the electron beam or as a compensation for the two upstream bunch compressors, the different electron beam phase evolution in ERL mode is distinguished with red and green mark.

As we known, the quality factor $Q_0$ could be utilized to describe the ratio of the energy gained to the power dissipated in one radio-frequency circle, higher $Q_0$ means more acceleration per unit of RF power and therefore determines the efficiency of the RF system in Linac. The quality factor with different beam loading is $Q_L$, the basic operation parameters for the whole module and RF station is calculated based on this value, the relationship between $Q_0$ and $Q_L$ could be described as: [25,26]

$$Q_L = Q_0/(1+\beta) \tag{1}$$

Where $\beta$ is the optimized coupling factor, it can be written as,

$$\beta = \sqrt{(b+1)^2 + [2Q_0\frac{\Delta f}{f_0} + b \times tan(\varphi_t)]^2} \tag{2}$$

$b = I_t R_a \cos(\varphi_t)/V_c$ , $I_t$ is the magnitude of the resultant beam current vector in the cavity, $R_a$ is the shunt impedance of cavity, $\varphi_t$ is the phase of the resultant beam vector with respect to the cavity voltage, and $\Delta f$ is the residual microphonics detuning. The RF power requirements for a single 1.3 GHz cavity can be presented as:

$$P_f(Q_L, \Delta f) = \frac{V_c^2}{4(R/Q)Q_L}[\left(1 + \frac{I_t}{V_c}\frac{R}{Q}Q_L cos(\varphi_t)\right)^2 + \left(2Q_L\frac{\Delta f}{f_0} + \frac{I_t}{V_c}\frac{R}{Q}Q_L sin(\varphi_t)\right)^2] \tag{3}$$

where $V_c$ is the accelerating gradient, $R/Q$ is the characteristic impendence and the value is 1036 ohm for a 1.3GHz cavity. Considering 10% overhead and 6% transmission losses, the minimum RF power requirement according to Eq. (1) - Eq. (3) is 6.8 kW in SHINE for 0.3 mA average beam current, 16 MV/m accelerating gradient and 10Hz detuning offset. Fundamental power coupler is another critical element that would affect the possibility of ERL operation in L4. Fundamental power coupler is utilized to transmit the power generated by power source to the cavity and determine the energy admission rate can be delivered to the electron beam, full reflection of the RF input power at any relative phase should be accommodated. We compare the reflected power ($P_t$) in coupler for ERL and normal operation with Eq. (4).

$$P_r = P_f - I_t V_c \cos(\varphi_t) \tag{4}$$

Fig. 2(a) shows the loaded quality factor evolution with the beam current in different cases. The attenuation can be clearly seen with the increase of average electron beam current, the minimum loaded quality factor for SHINE is about $4.1 \times 10^7$ in the nominal gradient of 16 MV/m and 0.3 mA average current. A higher accelerating gradient could improve the beam loaded quality factor, as showed by the blue curve, which means a more efficient energy transfer could be realized under the permission of RF station and related instruments. With the development of processing technologies such as electro-polishing, N-doping et.al. in superconducting cavities,[27-28] the effective gradient of a cavity could be improved to about 30 MV/m or even higher, with the $Q_0$ still maintained in a very high level.[29-31]

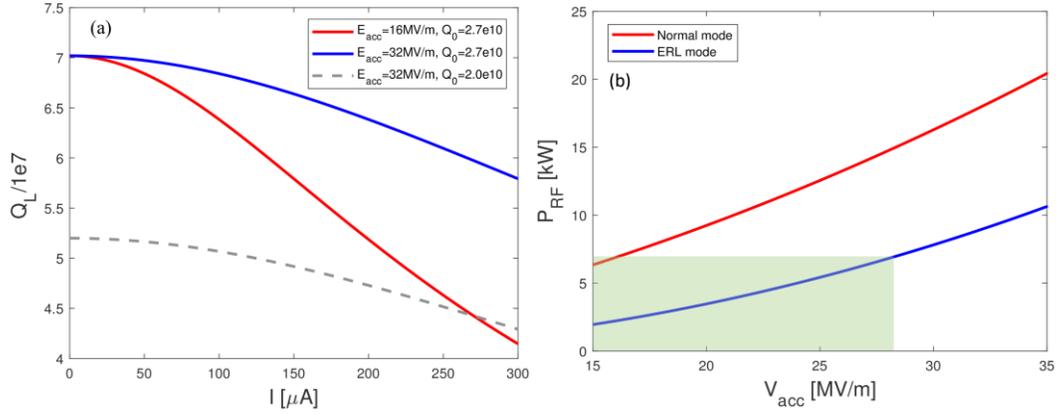

Fig. 2. The evolution of loaded quality factor with average current (left), and the comparison of RF power requirements for ERL and normal operation, the green area shows the feasibility accelerating gradient for the two different operating modes.

The capability of RF power is one of the major restrictions to prevent the cavities working in higher accelerating gradient. We compared the RF power requirements in ERL and normal operation in L4 section. The results for RF generator are summarized in Fig. 2(b). The power requirement is over 50% lower in ERL operation during the accelerating gradient range from 15-35 MV/m. For the nominal gradient of 16 MV/m, the 54 cryo-modules in L3 and L4 section could accelerate the electron beam from 2.1 to 8.74 GeV at full loaded, while the RF power requirements in L4 would be decreased to about 2.22 kW in ERL mode, which is only about 32 percent of the normal operation. The residue capacity of RF station enables a higher accelerating or decelerating gradient in L4, thus the energy of electron beam after the L3 could be adjusted continuously and widely. The capability of this new method is tradeoff between the number of cryo-modules and the performance limitation of the RF station and feasible accelerating gradient. When the capability of RF station is decided as 6.8 kW, the green area in Fig. 2(b) shows that the available accelerating gradient would be 16 MV/m and 27 MV/m for the two working modes, respectively. It is worth to emphasizing that, the 2K cryogenic power will be increased from 76 to 216 W for each cryo-module, and the overcapacity margin of SHINE cryogenic power will enable the operation of nearly 40 cryo-modules working in 27 MV/m. These results mean that, reasonable enhancement of the accelerating gradient in L4 will introduce about 69% maximum additional energy gain for the accelerated electron beam, without any upgrade or modification in RF station and cryogenic system.

What's more, the requirements in coupler is also considered for the actual operation status of SHINE. Because of the zero-beam loading in L4 during the ERL operation, all of the incident power is practically reflected back from the cavities. According to Eq. (4), it would be about 2.22 kW in 16 MV/m, while this value would be floating between 0.773 to 2.22 kW in normal operation, decided by the different beam loading. In full power ERL operation, the reflected power would be 6.8 kW in 27 MV/m, even though obviously stronger than normal mode, this value is still tolerable with the present coupler design due to the requirements to accommodate full input power reflection at any phase, and with the developments of more high average RF power couplers for superconducting cavities,[32,33] the capability of couplers to handle higher forward and reflected power could be further improved definitely.

In following, we discuss the optimal working point of ERL mode for SHINE case, i.e., the cryo-module numbers of L3 and L4. Considering the feasibility of delivering multi-energy electron beam to specific undulator line, the minimum energy of the decelerated beam is assumed to be no less than 0.5 GeV. The relationship between the cryo-module number of L4 and the maximum energy of the accelerated beam, and the corresponding accelerating gradient of L4 is shown in Fig. 3. In view of the performance of existing RF station, the optimized cryo-module number in L4 is decided as 25, the accelerating gradient is 26.26MV/m, and the energy of electron beam located at positive and negative phase is 11.41 GeV and 0.5 GeV, respectively.

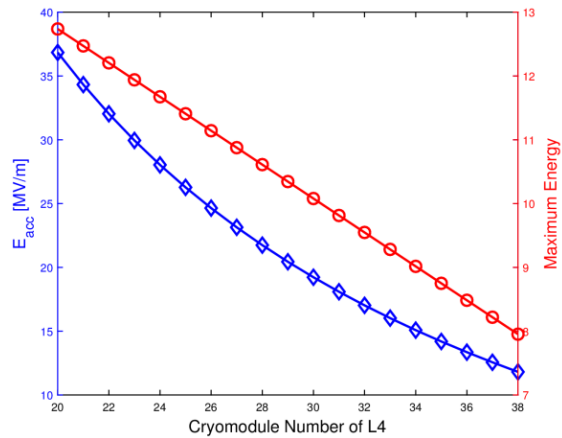

Fig. 3. The possible cryomodule number of L4 vs. the maximum energy output and corresponding accelerating gradient.

With the above parameters, start-to-end tracking of the electron beam, including all the components of SHINE, has been carried out. The electron beam dynamics in the photo-injector was simulated with ASTRA[34] to take into account space-charge effects. ELEGANT[35] was then used for the simulation in the remainder of the Linac, a two-stage compression is utilized in SHINE, and longitudinal beam phase space at the exit of the Linac is shown in Fig. 4. Due to the coherent synchrotron radiation (CSR) effect in the delay system, a little distortion happened in the head and tail of the electron beam, but the quality of electron beam central part is still maintained, the energy spread is about 9.5 MeV, the emittance is 0.2 μm-rad, the peak current is about 2kA, and the beam energy at exit of Linac is 0.5 and 11.41 GeV, respectively. The optimized beam lattice in L4 is also shown in Fig. 4, the reasonable TWISS parameter means that the transverse electron beam envelope and orbit is still acceptable in the accelerator. The flexibility of multi-energy operation has been experimentally demonstrated in SACLA.[36]

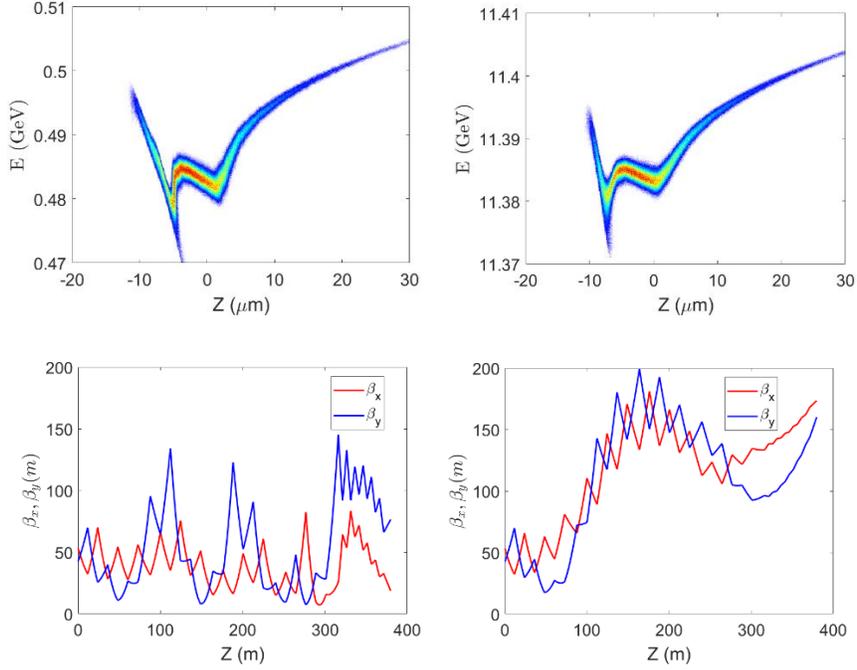

Fig. 4. The optimized longitudinal phase space (upper) and lattice (bottom) for 0.5 GeV (left) and 11.41 GeV (right) cases in L4 section.

For the FEL performances, we take the FEL-I case of SHINE as an example to illustrate the improvements by ERL operation. FEL-I covers the photon energy of 3-15 keV with a nominal 8.0 GeV electron beam. It will be operated in self-amplified spontaneous emission or self-seeding mode, and utilize the planar undulator with a period length of 26 mm.[5] Now with 11.41GeV electron beam at ERL mode, an FEL photon energy unreachable before, e.g., 30 keV, could be possible in FEL-I operation now. And a higher beam energy may significantly improve the FEL output performances when compared with the normal case. As a specific example, the self-seeding case at 12.94keV photon energy is simulated by the FEL code GENESIS[37] and 3D Bragg diffraction code BRIGHT.[38] In the simulation, undulator configuration of 7 cells (upstream) + 32 cells (downstream) is utilized, with a quadratic taper in the downstream undulator to enhance the FEL amplification efficiency. The optimized taper profile is characterized by a start-taper point of 15 m and a total taper of 2.0% over the length of the whole taper region. Fig. 5 shows the 12.94 keV FEL pulse energy evolution along the downstream undulator for 11.41 GeV electron beam. The maximum FEL output of 6 mJ is achieved, and the corresponding peak power exceeds 400 GW with a FWHM spectrum bandwidth of 0.9 eV. This unprecedented pulse energy could greatly ensure the successful development of exploration in vacuum birefringence, simulation of black hole physics, and generation of new forms of matter in SHINE.[39] At last, it's worth stressing that all the discussion here is based on the 0.3 mA average beam current, when the Linac is working with a lower beam loading, the tolerance of RF station will be greatly released, and the FEL tuning range can be future improved by increasing the maximum beam energy.

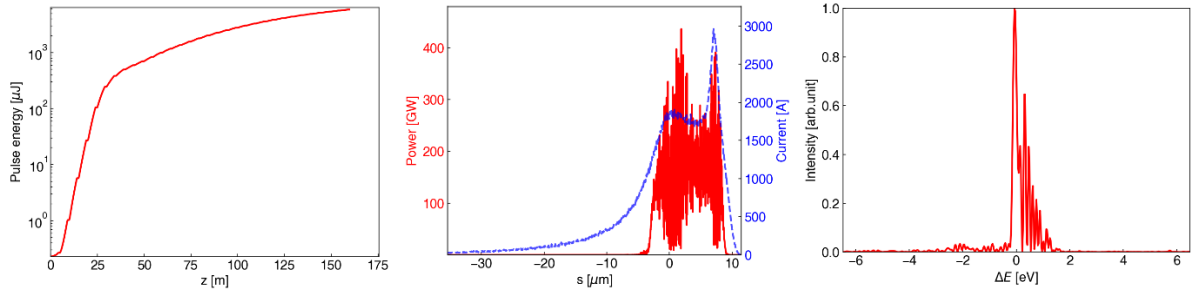

Fig. 5. The simulated self-seeding performance at 12.94keV photon energy of SHINE, with 11.41GeV beam energy at ERL mode.

In summary, the possibility of ERL operation in continuous-wave XFEL facility is investigated by using theoretical analysis and numerical simulations. It's found that this new method will greatly extend the tuning range of electron beam energy, approximately from 0.5 GeV to 11.41 GeV in 0.3 mA case of SHINE. The performance limit is mainly determined by the capability of RF station and the feasible accelerating gradient of superconducting cavities. Due to the net beam loading can be considered as zero in ERL operation, the challenges in the RF system will be greatly reduced, in other words, a higher gradient could be achieved in ERL operation with the original machine parameters. The reasonable electron beam control shows the feasibility of multi-energy beam transport with the same magnet parameters in the SRF Linac, and the deviations of the transverse beam envelop and beam parameters are still acceptable at the same time. By controlling the delay arc parameters, multi-energy beams could be adjusted continuously, and even exceed the nominal performance limit of SRF Linac. The proposed method in this Letter allows to improve the usability and enlarge the spectrum range of XFEL facility significantly.

The authors would like to thank Z. Y. Ma, H. T. Hou and Y. B. Zhao for helpful discussions on RF system. This work was partially supported by the National Key Research and Development Program of China (2018YFE0103100, 2016YFA0401900) and the National Natural Science Foundation of China (11935020, 11775293).

## DATA AVAILABILITY

The data that support the findings of this study are available from the corresponding author upon reasonable request.